\documentclass[12pt,preprint]{aastex}

\def\kms{\relax \ifmmode {\,\rm km\,s}^{-1}\else \,km\,s$^{-1}$\fi}
 
\def\degree{\mbox{$^{\circ}$}}
\def\Mso{{M$_{\rm \odot}$~}}
\def\mlr{\rm M$_\odot$~yr$^{-1}$}
\def\cm3{${\rm cm}^{-3}~$}

\def\oiii{[O {\sc iii}]}
\def\ha{H$\alpha$~}

\slugcomment{}

\shorttitle{PNe in the IC medium}
\shortauthors{Villaver \& Stanghellini}
  
\begin{document}
  
\title{The Survival of Planetary Nebulae in the Intracluster Medium}
\author{Eva Villaver\altaffilmark{1}}
\affil{Space Telescope Science Institute, 3700 San Martin Drive,
Baltimore, MD 21218, USA; villaver@stsci.edu}
\author{Letizia Stanghellini\altaffilmark{2}}
\affil{National Optical Astronomy Observatory, 950 N. Cherry Av.,
Tucson, AZ  85719, USA; lstanghellini@noao.edu}

\altaffiltext{2}{Affiliated with the Hubble Space Telescope Division
of the European Space Agency} 
\altaffiltext{3}{On leave from INAF-Bologna Observatory}  

\begin{abstract}

The stellar population stripped from galaxies in clusters evolve under  the
extreme conditions imposed by the intracluster (IC) medium. Intracluster
stars generally suffer very high systemic velocities, and
evolve within a rarefied and extremely hot IC medium. We present numerical 
simulations which aim to explore the evolution and survival of IC Asymptotic
Giant Branch (AGB) envelopes and Planetary Nebula (PN) shells. Our  
models reflect the evolution of a low-mass star under the observed conditions
in the Virgo IC medium. We find that the integrated
hydrogen-recombination line emission of a PN is
dominated by the inner dense shell, whose evolution is unaffected by
the environment. Ram pressure stripping affects mainly 
the outermost IC PN shell, which hardly influences the emission
when the PN is observed as a point source.  More importantly, 
we find that a PN with progenitor mass
of 1\Mso fades to $\sim$30\% and $\sim$10\% of its maximum emission, in
5\,000 and 10\,000 {\rm yr} respectively, disclosing 
an actual PN lifetime (t$_{PN}$) several times shorter to what is 
usually adopted (25\,000 {\rm yr}). This result affects the theoretical
calculation of the luminosity-specific density of IC PNe, which scales with
t$_{PN}$. For t$_{PN}$=10\,000 {\rm yr}, our more 
conservative estimate, we obtain that the luminosity-specific density of
PNe is in fair agreement with the value obtained
from Red Giants. With our more realistic PN 
lifetime we infer a higher fraction (above 15\%) of IC starlight in the
Virgo core than current estimates.
\end{abstract}


\section{INTRODUCTION}
In the current paradigm, the formation of large-scale structure in the
Universe takes place in a hierarchical mode and galaxy clusters are built
through mergers of smaller objects. In this scenario, galaxies in clusters
undergo fast encounters (``galaxy harassment'', \citealt{Moo:96}), and tidal
interactions (``tidal stirring'', \citealt{May:01}) that allow stars to
escape the gravitational  field of their original galaxy, building a diffuse
stellar population in the intracluster (IC) medium \citep{Dre:84}.  The total
mass, structure, and kinematics of this stripped stellar component can be
used to test the cluster's dynamical history.

The distribution of IC stars is expected to be either smooth or clumpy
depending on when in the 
cluster's lifetime the stellar component was removed \citep{Mer:84,Rm:83}.  
An unrelaxed IC stellar
component with significant substructure has been predicted by \cite{Nap:03} 
on the basis of a N-body cosmological simulation of a Virgo-like cluster.
More elaborated simulations of the IC stellar component by
\cite{Mur:04} and \cite{Som:05} predict that the fraction of the IC 
stars removed increases with the cluster mass.

From an observational point of view, the diffuse IC component is estimated
both from surface brightness measurements and from the detection of
individual IC stars.  In the last few years, quantitative estimates of the
diffuse IC stellar population have been inferred  from the study of Red
Giants and Asymptotic Giant Branch (AGB) stars in the Virgo cluster
\citep{Ftv:98,Detal:02}, Planetary Nebulae (PNe) in the Virgo and Fornax
clusters and the Leo I group
\citep{Metal:97,Fcj:98,Fcj:03,Cjf:98,Aetal:96,Aetal:02,Aetal:03,Cas:03}, 
Supernovae (SNe) type Ia in the Virgo cluster \citep{Smi:81,Gms:02,Getal:03},
and Novae in the Fornax cluster \citep{Netal:05}.

PNe, typically observed via their bright
\oiii~$\lambda$5007\AA~emission line, are among the 
most useful tracers of IC light to date \citep{Fel:03,Arn:04}. Photometric
detection is relatively straightforward through narrow-band filters and their
spectroscopic confirmation is possible using moderate resolution
spectra. Observations of PNe in the  Virgo cluster allow to estimate that the
IC stellar component is $\sim$10\%  of all stars in the cluster
\citep{Aetal:02,Aetal:03,Fetal:04}.

While IC PNe are now widely used as probes of the stellar populations
stripped from galaxies, and ultimately used to test the cluster's dynamical
history, their formation,  evolution, and mere survival in the IC environment
has never been analyzed in detail. The stellar population stripped  from
galaxies in clusters evolve under the extreme conditions imposed by the IC
medium.  Typically, the stars have very high systemic velocities (up to
2000~{\rm km~s}$^{-1}$) moving through a rarefied, extremely hot X-ray
emitting IC medium. These extreme conditions might affect the PN formation
process as well as their survival.

In this paper we present a first attempt to explore the formation and
evolution of PN within the IC environment. The idea is to test how the high
systemic velocity of the star and the high thermal pressure conditions of the
IC medium 
affect the evolution of the AGB circumstellar shell and  the survival of the
PN. The ultimate goal is to understand if what we know about PNe in our and
in nearby galaxies can be extrapolated to infer the properties of the PN
population stripped from galaxies in clusters. In Section $\S$2 we describe
the   numerical method and the initial and boundary conditions used in the
simulations. In $\S$3 we present our results and the relevance they bear in
the IC light studies. Finally, in $\S$4 we summarize our results and draw the
conclusions of this study.

\section{THE NUMERICAL MODEL: INITIAL AND BOUNDARY CONDITIONS}

Our numerical calculation follows the evolution of an AGB star of 1 \Mso
(Main Sequence mass) through the IC medium (with
the conditions observed in the Virgo IC medium) until it produces a PN. The
stellar wind changes as we 
follow the evolution from the   early-AGB, before the onset of the first
thermal pulse, to the post-AGB phase. The modulations of the mass-loss  and
wind velocity during the TP-AGB phase have been taken from \cite{Vw:93}. The PN
is formed later, when the ejecta is photoionized and shaped by the fast wind
from the hot central star. During the post-AGB phase the evolution of the
ionizing radiation field and wind are computed following the stellar
evolution in the Hertzsprung--Russell (HR) diagram (from
\citealt{Vw:94}). Details of the wind assumptions and gas evolution for
different initial configurations can be found in \cite{Vgm:02} and
\cite{Vmg:02}.

We assume that the star evolves while moving through the IC medium  with a
velocity of 1\,000 \kms, the typical PN velocity observed in the Virgo
IC medium \citep{Aetal:04}. The IC medium gas was assumed  to be homogeneous,
with a temperature of $T=10^7$ {\rm K} (4.75 {\rm keV}) \citep{Tak:89}, and a
density of $n=10^{-4}$ \cm3 \citep{Fg:83}. We consider that  the IC
pressure is simply  the standard gas thermal pressure, $P=n k T$ (where $k$
is the Boltzmann's constant).

We follow the evolution of the star through the IC medium using the ZEUS-3D
code (version 3.4), developed by  M.~L. Norman and the Laboratory for
Computational Astrophysics \citep{Sn:92a,Sn:92b,Smn:92}. This is a fully
tested code that solves the equations of hydrodynamics (when the magnetic
module is off) in a finite-difference, Eulerian, fully explicit scheme.  We
use radiative cooling which is made according to the cooling curves of
\cite{Rs:77}, and \cite{Dm:72} for gas temperatures above $10^4$ {\rm K}, and
according to \cite{Mb:81} for temperatures between $10^2$ {\rm K} and $10^4$
{\rm K}. 
ZEUS-3D does not include radiation transfer, so we use the approximation
implemented by \cite{Gf:96} to derive the location of the ionization front
(IF) for arbitrary density distributions (\citealt{Bty:79};
\citealt{Fttb:89}, 1990).  The position of the IF is determined by assuming
that ionization  equilibrium holds at all times, and that the ionization is
complete within the ionized sphere and zero outside. We apply this
formulation by assuming that the nebula is composed of hydrogen, that it is
optically thick in the Lyman continuum, and that the  ``on the spot''
approximation is valid. The unperturbed gas is treated
adiabatically. Finally, the photoionized gas is always kept at $10^4$ {\rm
K}, thus no cooling  curve is applied inside the photoionized region, unless
there is a shock.

The computations have been performed in 2D in a spherical polar grid ($\rho$,
$\theta$, and $\phi$) by assuming axisymmetry in the $\phi$ coordinate.
The angular $\theta$ coordinate in the simulations
ranges from 0 to 180${\degree}$ and the physical radial extension is 3 {\rm
pc} which gives us a total grid size of 6 x 3 {\rm pc$^2$}. We have opted for a
medium numerical resolution (400 x 360 zones in the radial and angular
coordinates of the grid respectively) to solve the problem. 

In order to maintain all the details of the
wind modulations suffered by the star we have imposed an additional restriction
to the Courant-Friedrichs-Levi condition to avoid the timestep of the
computation being larger than the timestep of stellar evolution.  

We set the time-dependent wind parameters (velocity, mass-loss rate, and wind
temperature) within a small spherical region (five radial zones) centered on
the symmetry axis where we used a reflecting boundary condition. The
density, velocity (normal and tangential components) and energy of the IC
medium are updated at each timestep to reflect the inflow that takes place in
the angular coordinate from 0 to 90$\degree$, and the outflow from 90 to
180$\degree$. As similar approach has been used in \cite{Vgm:03} to study the
evolution of a slowly moving star in the Galaxy.

\section{RESULTS}
\subsection{The Gas Evolution}
The models cover an evolutionary time of 5.3$\times$10$^5$ {\rm yr}, 93$\%$ of
which is spent on the  TP-AGB phase.  The computation starts at the onset of
the TP-AGB phase and is followed until the PN is 30\,000 {\rm yr} old. The
zero age for the PN phase 
is set when the star reaches a temperature of 10\,000 {\rm K} and thus emits
a non-negligible amount of ionizing photons. A transition time of 13\,000
{\rm yr} between the AGB and the PN phase has also been included in the
simulations. During the transition time we have assumed that the effective
temperature of the star increases linearly from the value at the tip of the
AGB to 10\,000 {\rm K}, which represents the starting temperature of the
post-AGB evolution. Note that we are
implicitly assuming that the removal of the star from the parent galaxy had
happened in the 10$^{10}$ {\rm yr} before the star reaches the TP-AGB.

In Figure~1 we show snapshots of the gas density (logarithm scale) at different
stages during the AGB evolution of the star. The star is fixed at the center
of the grid and the IC medium flows from top to bottom.  The computational
grid has been projected over the $\theta$ axis for illustration purposes. 
The first snapshot in the top left panel of Fig~1 represents the gas density at
2.8$\times$10$^5$ {\rm yr} from the start of the TP-AGB evolution. The
subsequent plots (from top to bottom, left to right) show the evolution at
time intervals of  24\,000 {\rm yr},  with the exception of the last 3 plots
choosen at time intervals of 12\,000 {\rm yr}.  

Prior to the so-called superwind phase \citep{Ren:81}, during the early
TP-AGB (the first $\sim$2.2$\times$10$^5$ {\rm yr}) the 
stellar mass-loss rate and velocity are small and constant with typical
values of a Red Giant wind (10$^{-8}$ \mlr~and 2\kms~respectively). Note that
this is before the first snapshot shown in Fig.~1. Ram
pressure balance is reached very close to 
the position of the star in the direction of the motion. The stagnation
radius (the radius at which pressure balance is reached for a free-streaming
wind) is 0.025 {\rm pc}, and a long stream of gas is generated in the
opposite direction to the motion. After this early stage, the
mass-loss rate and wind velocity are changing continuously at the inner
boundary as the star evolves up the TP-AGB.
The star undergoes four major thermal pulses of different duration in which
the mass-loss rate increases from 10$^{-8}$\mlr~to
5.3$\times$10$^{-4}$\mlr~and the wind velocity from 
2\kms to 15\kms (for details see \citealt{Vgm:02}). At any given time, this
continuosly evolving wind cannot be used to compute the stagnation radius,
because as shocks develop inside the bow-shock cavity formed by the early
time-independent wind, the wind cannot be considered to be free-flowing to
reach pressure balance with the IC medium. The ram pressure of the IC 
medium is balanced by the ram pressure of the stellar wind inside the
bow-shock cavity, making it a time-dependent problem. Note that the fluid is
supersonic for the relative velocity we are considering (1\,000 \kms) between
the star and the IC medium.

After 2.6$\times10^{5}$ {\rm yr}, the star undergoes the first major
thermal pulse and a bow-shock grows in size in the upstream direction.
This is the bow-shock visible at 2.8$\times10^{5}$ {\rm yr} shown in the top
left panel of Fig.~1. In the downstream direction, close to the symmetry
axis, the flow turns back. Long tongues of 
gas, consequence of the stripping process, are also visible. Note the
turbulent nature of the stripped gas. About 50\,000 {\rm yr} later (third
panel from the top left) the second major thermal pulse has already taken
place and subsequently the bow-shock feature in the upstream direction grows in
size. About 20\,000 yr later, an instability develops in the upstream
direction of the bow-shock at an opening angle of $\sim$ 30$\degree$ that
breaks the thin shell formed by the bow-shock.

In the lower panels of Fig.1, the density structure show the effects of the
increase in the  mass-loss rate taking place at the final thermal pulse. The
wind at this stage has enough momentum to allow the  formation of the external
shell 
that increasingly grows in size. As a consequence of the interaction with the
IC medium gas is continuously removed from the external shell. The material
left behind by the star generates a  long stream of gas that will have
a size of the order of 130 {\rm pc} if dispersion is  prevented.
 
\subsection{The Evolution of the PN Line Emission}
In order to gauge the observational differences between a PN evolving in the
IC medium and one within a galaxy we have computed the 
evolution of their \ha emission. Note that for the galactic PNe models we refer
to models in which there is no relative motion between the star and its
surrounding medium. In Figure~2 we show the evolution of the 
integrated \ha emission for the IC (dots) and a galactic (solid line, from
\citealt{Villa:01}) nebulae, each normalized to its own emission maximum.
The dashed line represents a galactic model without transition time evolution
\citep{Vmg:02}. The \ha intensity in each case has been computed by
integrating the \ha emission coefficient over 
nebular volume, assuming an electron temperature of  10\,000 {\rm K}, and
case B for the recombination \citep{Ost:89}. 

As shown in Fig.~2 the \ha line emission decreases very rapidly in both the
IC and galactic models. The IC PN emission fades to $\sim$30\% of its maximum
in 5\,000 {\rm yr}, and to $\sim$10\% of its maximum in 10\,000 {\rm yr}. The
evolution of the intensity directly reflects the development of dense regions
associated with 
shocks. The first peak in intensity (solid and dashed lines in Fig.~2) is
caused by the propagation 
of the shock front associated with the ionization front. The intensity
maximum is caused by a combination of two effects: the compression generated in
the inner shell as the hot bubble develops when the  
wind velocity increases as the star evolves toward higher effective
temperatures, and the transition from an optically thick to an optically thin
nebula.  

The differences between the solid and dashed lines in Fig.~2 are due to the
evolution during the transition time (the time elapsed between the end of the
TP-AGB and the PN ionization). The time at which the \ha 
intensity maximum occurs depends on  the  density structure encountered by
the wind and radiation field during the post-AGB phase. During the transition
time, the circumstellar gas structure developed during the AGB is
diluted. As a consequence, when no transition time is allowed (dashed line),
the ionization and fast wind encounter a denser structure,  the propagation
velocity of the ionization front is retarded, and the transition from
optically thick to an optically thin nebula takes longer. Moreover, both the
over-pressure generated by the photoionization and the gas density are higher,
preventing the expansion of the hot bubble.

We do not find significant differences in the \ha decay intensity between
the PN in the IC  medium (dots) and the galactic model, at least during the
bright portion of their evolution. The PN evolving in the IC medium is
undergoing strong ram pressure stripping, occurring mainly in the
outer shell. The \ha intensity is dominated by  the brightness of the main
nebular shell, at least in the $\sim$5\,000 {\rm yr} after  the \ha maximum,
and thus it is not affected by ram pressure stripping. However,
after 15\,000 {\rm yr}, the presence of the outer shell has a significant
effect on the emitted intensity,  but at this point the PN is already below
detectability limits  when large distances are involved (as is the case for
the IC PNe). It is worth 
noting that the temporal resolution we have used in the IC PN simulation
($\sim$6\,000 {\rm yr}) is much smaller than the one used in the galactic
models ($\sim$250 {\rm yr}) and thus we cannot follow all the fine details of
the \ha~intensity evolution of the IC PN simulation.

It is usually assumed that the evolution of a PN in the Galaxy and in the IC
environment are similar. In this work we confirm the validity of
this assumption, at  least where the decline of the PN \ha intensity is
concerned. What we find, however, is that in only 10\,000  {\rm yr} the
nebular intensity has dropped by a factor of ten from the value at maximum,
similarly for IC and galactic PNe. The 
number of PN in a given stellar population is  usually  computed assuming
that the  PN lifetime is $\sim$25\,000 {\rm yr}
\citep{Metal:93,Fel:03,Agu:05}. The effects of  
a shorter PN lifetime are discussed in \S3.3.

The physical reason why the bright PN stage is shorter in our calculation
than what is typically assumed can be easily
illustrated by simple reasoning. Generally, it is assumed that the hydrogen
recombination 
luminosity is nearly proportional to the number of ionizing photons, a
sensible assumption in a constant density regime. Our model reproduce a
more realistic structure that takes into account the hydrodynamic evolution
of the gas.  Since I$_{H_{\alpha}}$ $\propto$ $\int_{V}^{} n_e n_p\,dV$
(where n$_e$ and n$_p$ represent the number density of electrons and protons
respectively), the bulk of the recombination line intensity is completely
dominated by the PN 
regions where the density is higher, such as the  regions associated  with
shock fronts. The IC PN, at least under the conditions explored for the Virgo
cluster, have a similar fading time as models computed under
typical ISM conditions for a galaxy, because the inner shell is not
affected by ram pressure stripping. A much higher ram pressure
would be required to change the evolution of the bright PN shell.

Note that the logarithmic transformation of the intensity  fading is similar
to the functional shape of the PN luminosity function (PNLF), and that a
factor of ten decrease in intensity translates into an increase of 2.5
magnitudes. 

The PN detections in the IC medium are based on the flux measured on the
collisionally excited \oiii~5007~\AA~line. Since we do not solve radiation
transfer, we cannot describe the evolution of the
\oiii~5007~\AA~line intensity from our simulations. However, we expect the
evolution of the 
\oiii~5007~\AA~line and \ha to be similar, except in the case of very low
metallicity \citep{Setal:03}.

\subsection{The IC PN Lifetime and its Implications on the IC Starlight}
If, as our simulations suggest, the PN lifetime is indeed shorter than
usually adopted, the derived fraction of IC starlight contributing to the total
cluster luminosity should be revised. 

In order to predict the number of PNe in a given environment we  
can refer to the fuel consumption theorem \citep{Rb:86}, which states that
the number of post-main sequence (PMS) stars in a simple stellar
population (SSP) is proportional to the population luminosity and to the time  
elapsed in that particular PMS phase. Obviously, this theorem can be extended
to PNe if we rewrite it for the central star (CS) phase of
stellar evolution. By definition, t$_{\rm CS}$=t$_{\rm PN}$, thus, 
$N_{\rm PN}~=~B~L_{\rm T}~t_{\rm PN}$, where B is the evolutionary flux per
unit luminosity of the parent population 
and L$_T$ is the total bolometric luminosity of the sampled population. Our
model has a turnoff mass  of 1 M$_{\odot}$, thus B $\approx$
2$\times$10$^{-11}$ PN {\rm yr$^{-1}$} L$_{\sun}^{-1}$ \citep{Mar:98}.  The
theoretical luminosity-specific PN density can be expressed as
$\alpha$=N$_{\rm PN}$/L$_{\rm T}$.   If we assume that t$_{\rm PN}$=10\,000
{\rm yr} we find $\alpha$=2.0 $\times$ 10$^{-7}$ PN L$_{\odot}^{-1}$, while
the luminosity-specific population of the bright 2.5 magnitude bins in the
PNLF is about one tenth of this value, if we assume 
that the  PNLF has the shape given by \cite{Fcj:98}.  Our value of
$\alpha_{2.5}$=2$\times$10$^{-8}$ PN L$_{\odot}^{-1}$ agrees with
the observed value obtained from IC red giants \citep{Detal:02}.

In order to assess the effect of our IC PN lifetime estimate, we  evaluate
the luminosity-specific PN density in the brightest 1.0 magnitude bin of the
PNLF, as used by \citet{Agu:05}. From our lifetime estimate we find that
$\alpha_{1.0}$ is between 2.4 and 4.8 $\times$10$^{-9}$ PN L$_{\odot}^{-1}$
depending on whether we assume that the IC PN fades in  5\,000 or 10\,000 {\rm
yr}. For the longer fading time the stars in the Virgo core IC contribute
about 7$\%$ to the total starlight, as derived by \citet{Agu:05} using
$\alpha_{1.0}$ from \citet{Detal:02}. We are inclined  to believe that a
better estimate of the duration of the IC PN life is 5\,000 {\rm yr}, thus
the fraction of the IC starlight in the Virgo core would rise to $\sim$15$\%$
of the total starlight.

\section{CONCLUSIONS}
PNe have proved to be among the most useful tracers of IC light to
date. Nonetheless, no 
attempt has been 
made to date to explore the effects of the IC environment on their
evolution. We have simulated the evolution of a 1 \Mso star  
during the AGB and post-AGB phases, as it moves with the typical
velocity of 1\,000 \kms as is observed in the Virgo IC medium. 
We found that the evolution of such a star in the IC medium is different from
a galactic model in the details of the outer shell structure, but they have 
similar observable properties when studied as point sources. We based this
conclusion on the hydrogen-recombination integrated line emission,
and, ultimately, on the fact that the emission is dominated by the 
bright, inner shell of the PN, which evolves similarly in the IC and
galactic cases because it does not get affected by ram
pressure stripping until very late in the PN evolution.

These results, while valid to describe the evolution of a PNe in the rarefied
IC medium, are not adequate to describe a fast moving star through a much
higher density environment, such as that of the Leo I cloud, where no IC
PNe have been found \citep{Cas:03}.  We believe that a higher  ram pressure
might remove more 
material from the inner PN shell, reducing the IC PN lifetime much further,
thus preventing IC PN detection.

We have shown that the PN lifetime is shorter than usually assumed,  and we
believe that this apply to most IC and galactic environments. In the case
under study, 
the 1 \Mso star represent the turnoff mass of  a SSP approximately  10
{\rm Gyr}, and a fading time of 5\,000 {\rm yr} to 10\,000 {\rm yr} is
appropriate to 
theoretically predict the number of PNe per total luminosity.  We derived
that  $\alpha_{2.5}$=2$\times$10$^{-8}$ PN L$_{\odot}^{-1}$, which is in good
agreement with the observations of AGB stars.  By using our value for
$\alpha_{2.5}$ we 
inferred that the fraction of IC starlight to the total stellar luminosity
derived by \citet{Agu:05} and \cite{Fetal:04} are lower limits. We
estimate that the fraction of IC stellar light is greater than 10\% of the
total stellar luminosity for a SSP which  
is 10 {\rm Gyr} old.

\acknowledgments
We thank M. L. Norman and the Laboratory for Computational Astrophysics for
the use of ZEUS-3D.

\figcaption{
Snapshots of the (logarithmic) gas density generated during the AGB evolution
of a 1\,M$_{\odot}$ star moving in the IC medium. We 
have assumed a relative velocity of 1\,000 km\ s$^{-1}$, with the IC 
medium flowing into the grid from top to bottom. The evolution of the
stellar wind during the thermal-pulsing AGB phase 
and the PN stage has been implemented at the center
of the grid. The star suffers four major thermal pulses at 2.6$\times10^5$,
3.3$\times10^5$, 4.1$\times10^5$, and 4.8$\times10^5$ {\rm 
  yr}, each of them with a different
duration. From top to bottom and left to
right the snaphots have been taken at 2.8$\times10^5$, 3.0$\times10^5$,
3.3$\times10^5$, 3.5$\times10^5$, 3.8$\times10^5$, 3.9$\times10^5$,
4.0$\times10^5$, and 4.14$\times10^5$ {\rm yr} respectively.} 

\begin{figure}
\plotone{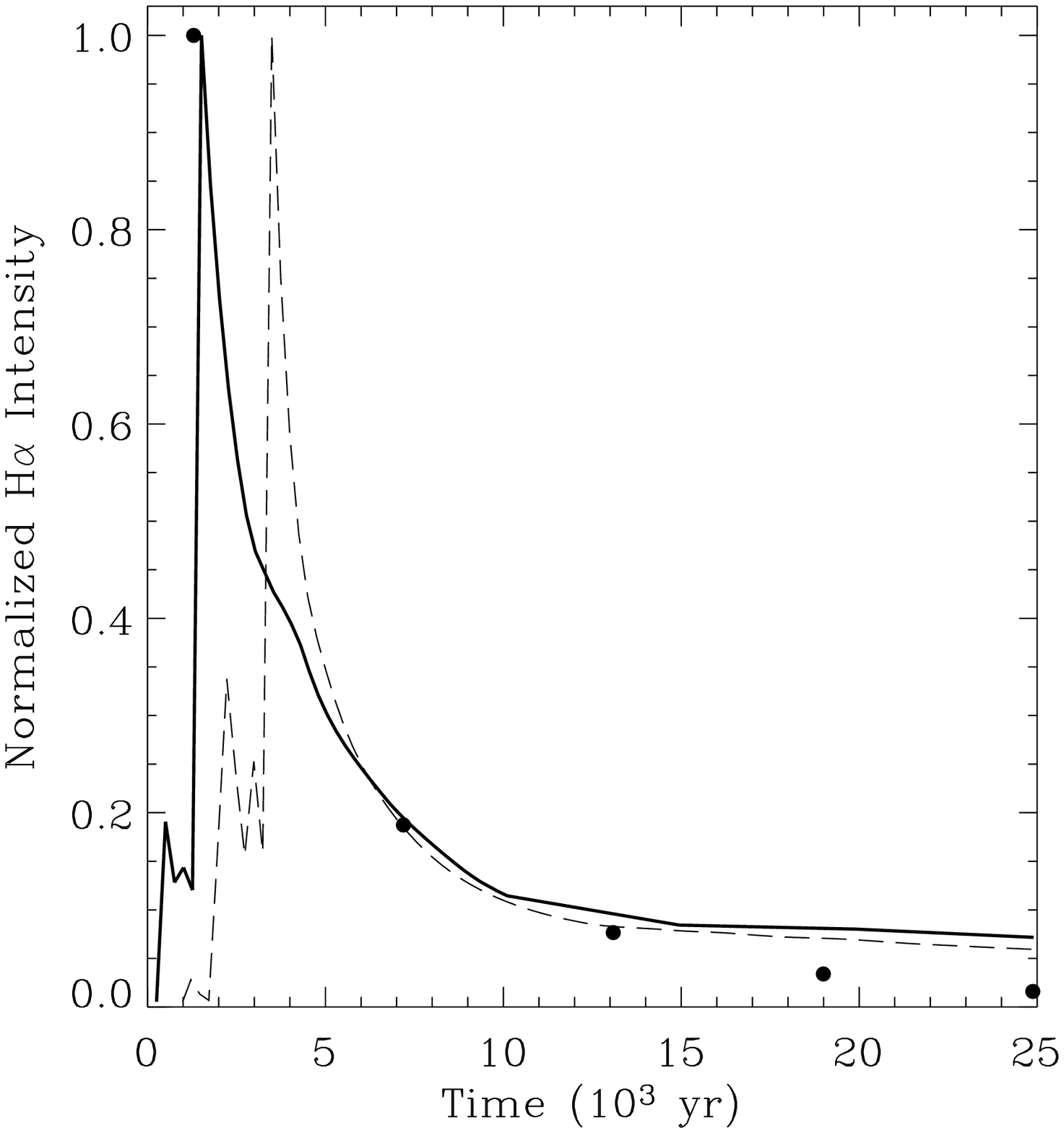}
\caption[ ]{The evolution of the total integrated \ha~intensity emitted by the 
PNe evolving in the Galaxy (solid line) and the IC medium (dots), normalized
to their maxima. In this figure the timescale represent the time   
elapsed after the PN illumination (i.e., when T$_{\rm eff}$=10\,000 {\rm K}).
Both PN models (Galaxy and IC medium) include evolution during the transition 
time (see text). As a comparison, a galactic model
without the evolution during the transition time is plotted as a dashed line.
\label{f2.eps}} 
\end{figure}


\begin{thebibliography}{}

\bibitem[Aguerri et al.(2005)]{Agu:05} Aguerri, J.~A.~L., 
Gerhard, O.~E., Arnaboldi, M., Napolitano, N.~R., Castro-Rodriguez, N., \& 
Freeman, K.~C.\ 2005, \aj, 129, 2585 

\bibitem[Arnaboldi(2004)]{Arn:04} Arnaboldi, M.\ 2004, IAU 
Symposium, 217, 5

\bibitem[Arnaboldi et al.(1996)]{Aetal:96}
Arnaboldi, M., Freeman, K. C., Mendez, R. H., et al. 1996, \apj, 472, 145   

\bibitem[Arnaboldi et al.(2004)]{Aetal:04} Arnaboldi, M., 
Gerhard, O., Aguerri, J.~A.~L., Freeman, K.~C., Napolitano, N.~R., Okamura, 
S., \& Yasuda, N.\ 2004, \apjl, 614, L33 

\bibitem[Arnaboldi et al.(2002)]{Aetal:02} Arnaboldi, M., et 
al.\ 2002, \aj, 123, 760 
\bibitem[Arnaboldi et al.(2003)]{Aetal:03} Arnaboldi, M., et 
al.\ 2003, \aj, 125, 514 


\bibitem[Bahcall et al.(2000)]{Betal:00}
Bahcall, N. A., Cen, R., Davé, R., Ostriker, J. P., \& Yu, Q. 2000, \apj,
541, 1  
\bibitem[Bodenheimer, Tenorio-Tagle, \& Yorke(1979)]{Bty:79}
Bodenheimer, G., Tenorio-Tagle, G., \& Yorke, H. W. 1979, \apj, 233, 85

\bibitem[Buzzoni(1989)]{Buz:89} Buzzoni, A.\ 1989, \apjs, 71, 
817 
\bibitem[Carlberg etal.(1996)]{Cetal:96}
Carlberg, R. G., Yee, H. K. C., Ellingson, E., et al. 1996, ApJ, 462, 32   
 
\bibitem[Castro-Rodr{\'{\i}}guez et al.(2003)]{Cas:03} 
Castro-Rodr{\'{\i}}guez, N., Aguerri, J.~A.~L., Arnaboldi, M., Gerhard, O., 
Freeman, K.~C., Napolitano, N.~R., \& Capaccioli, M.\ 2003, \aap, 405, 803 

\bibitem[Ciardullo, Jacoby, Feldmeier, \& 
Bartlett(1998)]{Cjf:98} Ciardullo, R., Jacoby, G.~H., 
Feldmeier, J.~J., \& Bartlett, R.~E.\ 1998, \apj, 492, 62 

\bibitem[Dalgarno \& McCray(1972)]{Dm:72}
Dalgarno, A. \& McCray, R.A. 1972, ARA\&A, 10, 375

\bibitem[Dressler(1984)]{Dre:84} Dressler, A.\ 1984, \araa, 
22, 185 

\bibitem[Durrell et al.(2002)]{Detal:02} Durrell, P.~R., 
Ciardullo, R., Feldmeier, J.~J., Jacoby, G.~H., \& Sigurdsson, S.\ 2002, 
\apj, 570, 119 

\bibitem[Fabricant \& Gorenstein(1983)]{Fg:83} Fabricant, D., 
\& Gorenstein, P.\ 1983, \apj, 267, 535 

\bibitem[Feldmeier(2003)]{Fel:03} Feldmeier, J.~J.\ 2003, IAU 
Symposium, 209, 597 

\bibitem[Feldmeier, Ciardullo, \& Jacoby(1998)]{Fcj:98} 
Feldmeier, J.~J., Ciardullo, R., \& Jacoby, G.~H.\ 1998, \apj, 503, 109 

\bibitem[Feldmeier, Ciardullo, Jacoby, \& 
Durrell(2003)]{Fcj:03} Feldmeier, J.~J., Ciardullo, R., 
Jacoby, G.~H., \& Durrell, P.~R.\ 2003, \apjs, 145, 65 

\bibitem[Feldmeier et al.(2004)]{Fetal:04} Feldmeier, J.~J., 
Ciardullo, R., Jacoby, G.~H., \& Durrell, P.~R.\ 2004, \apj, 615, 196 


\bibitem[Ferguson, Tanvir, \& von Hippel(1998)]{Ftv:98} 
Ferguson, H.~C., Tanvir, N.~R., \& von Hippel, T.\ 1998, \nat, 391, 461 

\bibitem[Franco, Tenorio-Tagle, \& Bodenheimer(1989)]{Fttb:89}
Franco, J., Tenorio-Tagle, G., \& Bodenheimer, P. 1989, Rev. Mexicana
Astron. Astrofis., 18, 65

\bibitem[Franco, Tenorio-Tagle, \& Bodenheimer(1990)]{Fttb:90}
Franco, J., Tenorio-Tagle, G., \& Bodenheimer, P. 1990, \apj, 349, 126


\bibitem[Gal-Yam, Maoz, Guhathakurta, \& 
Filippenko(2003)]{Getal:03} Gal-Yam, A., Maoz, D., Guhathakurta, 
P., \& Filippenko, A.~V.\ 2003, \aj, 125, 1087 

\bibitem[Gal-Yam, Maoz, \& Sharon(2002)]{Gms:02} Gal-Yam, A., 
Maoz, D., \& Sharon, K.\ 2002, \mnras, 332, 37 


\bibitem[Garc\'{\i}a-Segura \& Franco(1996)]{Gf:96}
Garc\'{\i}a-Segura, G. \& Franco, J. 1996, \apj, 469, 171

\bibitem[MacDonald \& Bailey(1981)]{Mb:81}
MacDonald, J. \& Bailey, M. E. 1981, \mnras, 197, 995

\bibitem[Maraston(1998)]{Mar:98} Maraston, C.\ 1998, \mnras, 
300, 872 

\bibitem[Mayer et al.(2001)]{May:01} Mayer, L., Governato, F., 
Colpi, M., Moore, B., Quinn, T., Wadsley, J., Stadel, J., \& Lake, G.\ 
2001, \apjl, 547, L123 
 
\bibitem[Mendez, et al.(1997)]{Metal:97}
Mendez, R. H., Guerrero, M. A., Freeman, K. C., et al. 1997, \apj, 491, L23   


\bibitem[Mendez et al.(1993)]{Metal:93} Mendez, R.~H., 
Kudritzki, R.~P., Ciardullo, R., \& Jacoby, G.~H.\ 1993, \aap, 275, 534 

\bibitem[Merritt(1984)]{Mer:84} Merritt, D.\ 1984, \apj, 276, 
26 

\bibitem[Moore et al.(1996)]{Moo:96} Moore, B., Katz, N., 
Lake, G., Dressler, A., \& Oemler, A.\ 1996, \nat, 379, 613 

\bibitem[Murante et al.(2004)]{Mur:04} Murante, G., et al.\ 
2004, \apjl, 607, L83 

\bibitem[Napolitano et al.(2003)]{Nap:03} Napolitano, N.~R., 
et al.\ 2003, \apj, 594, 172 

\bibitem[Neill et al.(2005)]{Netal:05} Neill, J.~D., Shara, 
M.~M., \& Oegerle, W.~R.\ 2005, \apj, 618, 692

\bibitem[Osterbrock(1989)]{Ost:89} Osterbrock, D.~E.\ 1989, 
Research supported by the University of California, John Simon Guggenheim 
Memorial Foundation, University of Minnesota, et al.~Mill Valley, CA, 
University Science Books, 1989
 
\bibitem[Raymond \& Smith(1977)]{Rs:77}
Raymond, J.C. \& Smith, B.W. 1977, ApJS, 35, 419

\bibitem[Renzini(1981)]{Ren:81} Renzini, A.\ 1981 in Physical Processes in
  Red Giants, ed.. I. Iben \& A. Renzini(Dordrech:Reidel), 431
 
\bibitem[Renzini \& Buzzoni(1986)]{Rb:86} Renzini, A., \& 
Buzzoni, A.\ 1986, ASSL Vol.~122: Spectral Evolution of Galaxies, 195 

\bibitem[Richstone \& Malumuth(1983)]{Rm:83} Richstone, 
D.~O., \& Malumuth, E.~M.\ 1983, \apj, 268, 30 

\bibitem[Smith(1981)]{Smi:81} Smith, H.~A.\ 1981, \aj, 86, 998 

\bibitem[Sommer-Larsen et al.(2005)]{Som:05} Sommer-Larsen, 
J., Romeo, A.~D., \& Portinari, L.\ 2005, \mnras, 357, 478 

\bibitem[Stanghellini et al.(2003)]{Setal:03} Stanghellini, L., 
Shaw, R.~A., Balick, B., Mutchler, M., Blades, J.~C., \& Villaver, E.\ 
2003, \apj, 596, 997 

\bibitem[Stone, Mihalas, \& Norman(1992)]{Smn:92}
Stone, J.M., Mihalas, D., \& Norman, M. L. 1992, ApJS, 80, 819

\bibitem[Stone \& Norman(1992a)]{Sn:92a}
Stone, J.M. \& Norman, M. L. 1992a, ApJS, 80, 753
\bibitem[Stone \& Norman(1992b)]{Sn:92b}
Stone, J.M. \& Norman, M. L. 1992b, ApJS, 80, 791
\bibitem[Takano et al.(1989)]{Tak:89} Takano, S., Awaki, H., 
Koyama, K., Kunieda, H., \& Tawara, Y.\ 1989, \nat, 340, 289 

\bibitem[Vassiliadis \& Wood(1993)]{Vw:93}
Vassiliadis, E. \& Wood, P. 1993, \apj, 413, 641

\bibitem[Vassiliadis \& Wood(1994)]{Vw:94}
Vassiliadis, E. \& Wood, P. 1994, \apj, 92, 125

\bibitem[Villaver(2001)]{Villa:01} Villaver, E.\ 2001, 
Ph.D.~Thesis, University of La Laguna
 
\bibitem[Villaver, Garc\'{\i}a-Segura, \& Manchado(2002a)]{Vgm:02}
Villaver, E., Garc\'{\i}a-Segura, G., \& Manchado, A. 2002a, \apj, 571, 880

\bibitem[Villaver, Garc{\'{\i}}a-Segura, \& 
Manchado(2003)]{Vgm:03} Villaver, E., Garc{\'{\i}}a-Segura, 
G., \& Manchado, A.\ 2003, \apjl, 585, L49 

\bibitem[Villaver, Manchado \& Garc\'{\i}a-Segura(2002b)]{Vmg:02}
Villaver, E., Manchado, A., \& Garc\'{\i}a-Segura, G. 2002b, \apj, 581, 1204


\end{thebibliography}
\end{document}